\documentclass[aps,showpacs,prb,reprint,groupedaddress,longbibliography]{revtex4-1}

\usepackage[colorlinks=true,linkcolor=blue,citecolor=blue,urlcolor=blue]{hyperref}

\usepackage{hyperref}
\usepackage{setspace} 
\usepackage{graphicx}
\usepackage{amsmath}
\usepackage{color}
\usepackage{amsmath}
\usepackage{amssymb}
\usepackage{verbatim}
\usepackage{latexsym}
\usepackage{enumerate} 
\usepackage{bm} 

\newcommand{\argmin}{\operatornamewithlimits{arg\,min}}

\begin{document}

\title{Vibrational averages along thermal lines}

\author{Bartomeu Monserrat}
\email{bm418@cam.ac.uk}
\affiliation{Department of Physics and Astronomy, Rutgers University,
  Piscataway, New Jersey 08854-8019, USA}
\affiliation{TCM Group, Cavendish Laboratory, University of Cambridge,
  J.\ J.\ Thomson Avenue, Cambridge CB3 0HE, United Kingdom}

\date{\today}

\begin{abstract}
A method is proposed for the calculation of vibrational quantum and thermal expectation values of physical properties from first principles. \textit{Thermal lines} are introduced: these are lines in configuration space parametrized by temperature, such that the value of any physical property along them is approximately equal to the vibrational average of that property. The number of sampling points needed to explore the vibrational phase space is reduced by up to an order of magnitude when the full vibrational density is replaced by thermal lines.
Calculations of the vibrational averages of several properties and systems are reported, namely the internal energy and the electronic band gap of diamond and silicon, and the chemical shielding tensor of L-alanine. 
Thermal lines pave the way for complex calculations of vibrational averages, including large systems and methods beyond semi-local density functional theory. 
\end{abstract}

\pacs{31.15.A-, 63.70.+h, 71.38.-k, 71.15.Dx}

\maketitle

\section{Introduction}


First-principles quantum mechanical methods have been successfully used to calculate a wide variety of properties for a large number of systems. The vast majority of calculations rely on the static lattice approximation: atomic nuclei are fixed at their equilibrium positions, and only electron motion is considered. In reality atomic nuclei undergo quantum and thermal motion, which leads to a vibrational renormalization of the static lattice value of physical properties. For example, the temperature dependence of physical observables almost exclusively arises from nuclear motion in systems with a band gap.

Vibrational corrections have been calculated for electronic~\cite{0295-5075-10-6-011,nanotube_t_dependence,PhysRevLett.105.265501,cannuccia_ejpb,PhysRevB.87.144302,elph_Si_nano,helium,gonze_gw_elph,monserrat_elph_diamond_silicon,patrick_molecule_solid_2014,elph_topological_prl,elph_topological_prb,jcp_ponce_convergence,topological_elph_jhi}, magnetic~\cite{pickard_original_nmr_vibrational,md_quadrupolar_coupling,md_nmr_lee,pickard_nmr_vibrations_organic,nmr_force_fields,dracinsky_md_nmr,dracinsky_pimd_nmr,monserrat_nmr_tdep_original}, structural~\cite{PhysRevB.71.205214,PhysRevB.87.144302}, and optical~\cite{PhysRevLett.107.255501,phonon_assisted_optical_absorption,giustino_nat_comm,ph_assisted_optical_zacharias} properties. 
In these studies, the exploration of the vibrational phase space requires a large number of sampling points. As a consequence, vibrational correction calculations typically utilize electronic structure methods that are computationally inexpensive, mostly semi-local density functional theory (DFT)~\cite{PhysRev.136.B864,PhysRev.140.A1133,dft_rev_mod_phys,PhysRevLett.45.566,PhysRevB.23.5048,PhysRevLett.77.3865}. It would be desirable to calculate vibrational corrections using more accurate electronic structure methods, such as hybrid functional DFT~\cite{b3lyp,hybrid_band_gaps,pbe0,hse06_functional,hse06_functional_erratum}, many-body perturbation theory~\cite{original_gw_hedin,0034-4885-61-3-002}, or quantum Monte Carlo (QMC)~\cite{RevModPhys.73.33}, as semi-local DFT has several known deficiencies~\cite{PhysRevLett.56.2415,PhysRevB.37.10159,gonze_gw_elph}. However, the computational expense of methods beyond semi-local DFT makes their routine use in this context impossible in most cases. 
An additional limitation imposed by the large number of sampling points concerns the system sizes that can be explored: current methods are not appropriate for studying large non-periodic systems such as those relevant in nanoscale applications.

In this work I propose a method to calculate quantum and thermal averages of general physical properties from first principles at a small computational cost, irrespective of system size. As a result, many-atom systems can be studied, and the calculations could be combined with a sophisticated treatment of the electronic structure.
The rest of the paper is organised as follows. In Sec.~\ref{sec:formalism}, the methods used so far to calculate vibrational averages are reviewed first (Sec.~\ref{subsec:review}), followed by a presentation of the alternative formalism proposed here (Secs.~\ref{subsec:tl} and \ref{subsec:tl_mc}). The computational details are discussed in Sec.~\ref{sec:comput}. In Sec.~\ref{sec:results}, the results of the calculations performed to validate the method are presented. The calculations report vibrational averages of the energies and band structures of diamond and silicon, and of the chemical shielding tensor in L-alanine molecular crystals, exemplifying the wide applicability of the proposed method. The conclusions are drawn in Sec.~\ref{sec:conclusions}.

\section{Formalism}  \label{sec:formalism}

\subsection{Quantum and thermal averages} \label{subsec:review}

The harmonic nuclear vibrational Hamiltonian of a solid is~\cite{wallace,born,maradudin} 
\begin{equation}
\hat{\mathcal{H}}_{\mathrm{vib}}=\sum_{n,\mathbf{k}}-\frac{1}{2}\frac{\partial^2}{\partial q_{n\mathbf{k}}^2}+\frac{1}{2}\omega_{n\mathbf{k}}^2q_{n\mathbf{k}}^2, \label{eq:hamiltonian}
\end{equation}
where $q_{n\mathbf{k}}$ is a normal mode coordinate associated with vibrational Brillouin zone (BZ) wave vector $\mathbf{k}$ and branch $n$, and $\omega_{n\mathbf{k}}$ is the corresponding vibrational frequency. Atomic positions are described with points in configuration space labelled by the vector $\mathbf{q}$. For a solid containing $N$ atoms, $\mathbf{q}$ is a $3(N-1)$-dimensional vector with elements $\{q_{n\mathbf{k}}\}$. 
The Hamiltonian in Eq.~(\ref{eq:hamiltonian}) is separable, and can be solved analytically for each degree of freedom $(n,\mathbf{k})$. 
The nuclear eigenstates are:  
\begin{equation}
|\phi_M\rangle=\frac{1}{\sqrt{2^MM!}}\left(\frac{\omega}{\pi}\right)^{1/4}e^{-\omega q^2/2}H_M(\sqrt{\omega} q), \label{eq:eigenstate}
\end{equation}
with the associated energy spectrum:
\begin{equation}
E_M=\omega\left(\frac{1}{2}+M\right). \label{eq:eigenenergy}
\end{equation}
In Eq.~(\ref{eq:eigenstate}), $H_M$ is the Hermite polynomial of order $M$. 

Within the Born-Oppenheimer approximation~\cite{bornoppenheimer}, the harmonic expectation value of a physical property with associated observable $\hat{O}$ at temperature $T$ is
\begin{equation}
\langle \hat{O}(T)\rangle=\frac{1}{\mathcal{Z}}\sum_{\mathbf{M}}\langle\Phi_{\mathbf{M}}(\mathbf{q})|\hat{O}(\mathbf{q})|\Phi_{\mathbf{M}}(\mathbf{q})\rangle e^{-\frac{E_{\mathbf{M}}}{k_{\mathrm{B}}T}}, \label{eq:exp_val}
\end{equation}
where $|\Phi_{\mathbf{M}}\rangle=\prod_{n,\mathbf{k}}|\phi_{M_{n\mathbf{k}}}(q_{n\mathbf{k}})\rangle$ is a vibrational state of energy $E_{\mathbf{M}}$, $\mathcal{Z}=\sum_{\mathbf{M}} e^{-E_{\mathbf{M}}/k_{\mathrm{B}}T}$ is the partition function and $k_{\mathrm{B}}$ is the Boltzmann constant. $\langle \hat{O}(T)\rangle$ is also referred to as the vibrational average of observable $\hat{O}$. Note that the formulation of Eq.~(\ref{eq:exp_val}) does not include dynamical effects, which have been shown to be important in some systems~\cite{dynamical_anharmonic_cote}.

Equation~(\ref{eq:exp_val}) has been evaluated in the literature using three different families of methods: (i) the quadratic method, (ii) Monte Carlo methods, and (iii) dynamical methods. These approaches will be described next, before proposing an alternative that solves some of their limitations. 

\subsubsection{Quadratic method}

Consider the value of the property of interest at configuration $\mathbf{q}$ as an expansion about its value at the equilibrium configuration at $\mathbf{q=0}$,
\begin{equation}
\hat{O}(\mathbf{q})\!=\!\hat{O}(\mathbf{0})+\sum_{n,\mathbf{k}}a_{n\mathbf{k}}^{(1)}q_{n\mathbf{k}}+\!\!\sum_{n,\mathbf{k},n',\mathbf{k}'}\!\!\!\!a_{n\mathbf{k};n'\mathbf{k}'}^{(2)}q_{n\mathbf{k}}q_{n'\mathbf{k}'}+\cdots, \label{eq:expansion}
\end{equation}
where $\{a_{n\mathbf{k}}^{(1)},a_{n\mathbf{k};n'\mathbf{k}'}^{(2)},\ldots\}$ are the coupling constants, given by the change of the property under the corresponding atomic displacements. The expectation value becomes
\begin{equation}
\langle \hat{O}(T)\rangle=O(\mathbf{0})+\sum_{n,\mathbf{k}}\frac{a^{(2)}_{n\mathbf{k};n\mathbf{k}}}{2\omega_{n\mathbf{k}}}\left[1+2n_{\mathrm{B}}(\omega_{n\mathbf{k}},T)\right]+\mathcal{O}(q^4), \label{eq:quad_tdep}
\end{equation}
where $n_{\mathrm{B}}(\omega,T)=(e^{\omega/k_{\mathrm{B}}T}-1)^{-1}$ is a Bose-Einstein factor. Equation~(\ref{eq:quad_tdep}), referred to as the quadratic method, has been used to calculate vibrational averages of, amongst other properties, electronic band gaps (then known as Allen-Heine-Cardona theory~\cite{0022-3719-9-12-013}) and nuclear magnetic resonance parameters~\cite{monserrat_nmr_tdep_original}. Although this paper focuses on the harmonic approximation for the description of nuclear vibrations, note that the quadratic method has been extended to include anharmonic vibrations~\cite{PhysRevB.87.144302}. 

All $3(N-1)$ quadratic diagonal coupling constants $a^{(2)}_{n\mathbf{k};n\mathbf{k}}$ are needed for the evaluation of Eq.~(\ref{eq:quad_tdep}). These can be determined using frozen-phonon~\cite{Yin_1980,Fleszar_1985} or perturbative methods~\cite{Baroni_1987,PhysRevB.43.7231,Gonze_1997}, and symmetry reduces the number of explicit calculations required. Recent developments in the use of non-diagonal supercells have greatly expanded the applicability of the frozen-phonon method~\cite{non_diagonal}.



The quadratic method is the most widely used approach to calculate vibrational averages. This is because, compared to alternative methods, it typically requires the evaluation of the property of interest at the smallest number of configurations for systems containing up to a few hundred atoms. On the negative side, the number of sampling points required scales linearly with system size, and the use of the quadratic method for large systems is therefore limited. Furthermore, the higher order terms neglected in Eq.~(\ref{eq:quad_tdep}) have been shown to be important in some cases~\cite{helium,molec_crystals_elph,dynamical_anharmonic_cote}. 

\subsubsection{Monte Carlo methods}

The expectation value in Eq.~(\ref{eq:exp_val}) can be rewritten as: 
\begin{equation}
\langle \hat{O}(T)\rangle=\int\mathrm{d}\mathbf{q}|\Phi(\mathbf{q},T)|^2\hat{O}(\mathbf{q}), \label{eq:exp_val_mc}
\end{equation}
where $|\Phi(\mathbf{q},T)|^2$ 
is the nuclear density at temperature $T$, given by a product of Gaussian functions $(2\pi s^2)^{-1/2}\exp(-\frac{q^2}{2s^2})$, one for each degree of freedom $(n,\mathbf{k})$. The Gaussian widths $s$ depend on temperature:
\begin{equation}
s^2(T)=\frac{1}{2\omega}\coth\left(\frac{\omega}{2k_{\mathrm{B}}T}\right).
\end{equation}

Equation~(\ref{eq:exp_val_mc}) can be evaluated using Monte Carlo integration~\cite{pickard_nmr_vibrations_organic,giustino_nat_comm,helium}:
\begin{equation}
\langle \hat{O}(T)\rangle_{\mathrm{WF}}=\frac{1}{n}\sum_{i=1}^n\hat{O}(\mathbf{q}_i), \label{eq:mc}
\end{equation}
with sampling points distributed according to $|\Phi(\mathbf{q},T)|^2$, as indicated by the subscript WF for wave function. Monte Carlo integration is appropriate to evaluate high-dimensional integrals, as the number of sampling points required to achieve a certain statistical uncertainty in the result is independent of system size (if the variance of the integrand does not change with system size). Therefore, Monte Carlo integration will be the computationally cheaper approach beyond some system size. However, most published calculations consider systems containing up to a few hundred atoms, for which the number of sampling points is typically smaller using the quadratic method.

Monte Carlo integration is, in principle, more accurate than the quadratic method because it does not rely on the truncation of the expansion in Eq.~(\ref{eq:expansion}). If high-order terms are important, they are accurately captured by the Monte Carlo approach~\cite{helium,molec_crystals_elph}.

Finally, it should be noted that Monte Carlo sampling has been combined with anharmonic wave functions using a reweighting scheme~\cite{monserrat_nmr_tdep_original}.

\subsubsection{Dynamical methods}

Equilibrium properties of condensed phases can also be investigated using dynamical methods such as molecular dynamics or path integral molecular dynamics~\cite{md,ab_initio_md,pimd1,pimd2,PhysRevB.73.245202,PhysRevLett.101.106407,ceperley_h_elph_coupling,dracinsky_pimd_nmr,galli_water_gap_aimd}. Using these approaches, the system is evolved in time following classical or quantum dynamics, and the vibrational expectation value is calculated by averaging the value of the property of interest along the dynamical path. 

Dynamical methods capture anharmonic effects, and can be used to describe properties out of equilibrium. However, for the calculation of equilibrium properties, for which alternative methods exist, dynamical calculations tend to be computationally more intensive. This is caused by the intrinsic expense of calculating the dynamics of the system, and by the long paths typically needed to reduce statistical noise -- for example to remove serial correlation between samples.

\subsection{Thermal lines} \label{subsec:tl}

The methods used for the evaluation of vibrational averages complement each other in terms of accuracy and computational expense. However, an accurate \textit{and} computationally inexpensive method, though desirable, is currently not available. 
In the rest of this paper I present an alternative method to calculate vibrational averages $\langle \hat{O}(T)\rangle$ that satisfies both criteria. 

An atomic configuration $\mathbf{q}^{\mathrm{MV}}$ is sought for which the value of the observable is identical to the value of its vibrational average, 
\begin{equation}
\hat{O}(\mathbf{q}^{\mathrm{MV}})=\langle \hat{O}(T)\rangle, \label{eq:mean_value}
\end{equation} 
as dictated by the mean-value (MV) theorem for integrals. There is no general solution to Eq.~(\ref{eq:mean_value}), but under some suitable approximations it is possible to determine a good approximation to the mean-value configuration $\mathbf{q}^{\mathrm{MV}}$.

The quadratic diagonal coupling constants $a^{(2)}_{n\mathbf{k};n\mathbf{k}}$ in the expansion of Eq.~(\ref{eq:expansion}) are the only coupling constants below fourth order appearing in the vibrational average of Eq.~(\ref{eq:quad_tdep}). Based on this observation, and as a starting point, one can make the assumption that $a^{(2)}_{n\mathbf{k};n\mathbf{k}}$ are the dominant coupling constants in the expansion of Eq.~(\ref{eq:expansion}). One may then approximate
\begin{equation}
\hat{O}(\mathbf{q})\simeq\sum_{n,\mathbf{k}}a^{(2)}_{n\mathbf{k};n\mathbf{k}}q_{n\mathbf{k}}^2. \label{eq:quad_approx}
\end{equation}
From now on, the static lattice value $\hat{O}(\mathbf{0})$ is dropped from all equations. 
In the particular case that the observable in Eq.~(\ref{eq:quad_approx}) is the potential energy, then $a^{(2)}_{n\mathbf{k};n\mathbf{k}}=\frac{1}{2}\omega_{n\mathbf{k}}^2$ recovers the harmonic approximation. 

Using the expressions in Eqs.~(\ref{eq:quad_tdep}) and (\ref{eq:quad_approx}), which are referred to as the quadratic approximation (QA), it is possible to solve $\hat{O}(\mathbf{q}^{\mathrm{QA}})=\langle \hat{O}(T)\rangle$. A set of independent $3(N-1)$ equations determining $\mathbf{q}^{\mathrm{QA}}$ are obtained. Each of these equations corresponds to a degree of freedom $(n,\mathbf{k})$, with two possible solutions 
\begin{equation}
q_{n\mathbf{k}}^{\mathrm{QA}}(T)=\pm\left(\frac{1}{2\omega_{n\mathbf{k}}}\left[1+2n_{\mathrm{B}}(\omega_{n\mathbf{k}},T)\right]\right)^{1/2}. \label{eq:thermal_lines}
\end{equation}
The amplitude of the normal mode coordinates in Eq.~(\ref{eq:thermal_lines}) is equal to $\sqrt{\langle q_{n\mathbf{k}}^2\rangle}$. 

Equation~(\ref{eq:thermal_lines}) is the first central result of the paper. It provides an approximation to the mean value position, $\mathbf{q}^{\mathrm{MV}}\simeq\mathbf{q}^{\mathrm{QA}}$.
Each normal mode has two possible values in Eq.~(\ref{eq:thermal_lines}), and therefore there are $2^{3(N-1)}$ atomic configurations that solve Eq.~(\ref{eq:mean_value}) at each temperature $T$. A particular solution is characterized by the choice of positive or negative sign in each normal coordinate, and these signs define a $3(N-1)$-dimensional vector $\mathbf{S}$ with elements $S_{n\mathbf{k}}=\pm1$. The variation of temperature $T$ leads to a set of $2^{3(N-1)}$ lines in configuration space parametrized by $T$. I refer to these lines as \textit{thermal lines} $\mathcal{T}$:
\begin{equation}
\mathcal{T}_{\mathbf{S}}(T)\!=\!(S_1q_1^{\mathrm{QA}}(T),S_2q_2^{\mathrm{QA}}(T),\ldots,S_{3(N\!-\!1)}q_{3(N\!-\!1)}^{\mathrm{QA}}(T)).
\end{equation}
I refer to the starting points of thermal lines as \textit{quantum points}, corresponding to $T=0$ and with positions $q_{n\mathbf{k}}^{\mathrm{QA}}(T=0)=\pm1/\sqrt{2\omega_{n\mathbf{k}}}$. 

If Eq.~(\ref{eq:quad_approx}) were an exact equality, then the vibrational average of a physical property at temperature $T$ would be equal to the value of that physical property on any of the thermal lines at position $T$, $\langle \hat{O}(T)\rangle=\hat{O}[\mathcal{T}_{\mathbf{S}}(T)]$. Under such circumstances, any single thermal line would be sufficient to evaluate the vibrational average of a physical quantity according to Eq.~(\ref{eq:quad_tdep}). 


In reality, Eq.~(\ref{eq:quad_approx}) is not an exact equality. Averaging the value of the observable over pairs of opposite thermal lines:
\begin{equation}
\frac{1}{2}(\hat{O}[\mathcal{T}_{\mathbf{S}}(T)]+\hat{O}[\mathcal{T}_{-\mathbf{S}}(T)]), \label{eq:odd_term_suppression}
\end{equation}
removes all odd powers in the expansion of Eq.~(\ref{eq:expansion}), as each degree of freedom contributes with opposite signs. Therefore, by considering two, rather than one, thermal lines, a large part of the uncertainty associated with the approximation in Eq.~(\ref{eq:quad_approx}) is exactly removed. 

The remaining terms below fourth order are the off-diagonal quadratic terms, which can also be cancelled with an appropriate average over thermal lines. However, in this case the average is over $3(N-1)$ thermal lines. This is as expected, because the number of degrees of freedom in the system is the same as in the case of the quadratic approximation. Therefore, such averaging over thermal lines is an alternative approach to the evaluation of Eq.~(\ref{eq:quad_tdep}), but it requires the same number of points as the quadratic approximation, and therefore it provides no advantage. In particular, the scaling of the number of sampling points as a function of system size is still linear.

In the next section, a scheme is proposed to use thermal lines for the calculation of vibrational averages with a small, size-independent number of sampling points.


\subsection{Monte Carlo sampling over thermal lines} \label{subsec:tl_mc}

For atomic configurations along thermal lines, all vibrational normal modes contribute with amplitude $\sqrt{\langle q_{n\mathbf{k}}^2\rangle}$. This means that, although Eq.~(\ref{eq:quad_approx}) is not an exact equality, $\langle \hat{O}(T)\rangle\simeq\hat{O}[\mathcal{T}_{\mathbf{S}}(T)]$ is true for any thermal line $\mathbf{S}$. Furthermore, $\sum_{\mathbf{S}}\hat{O}[\mathcal{T}_{\mathbf{S}}(T)]=\langle\hat{O}(T)\rangle$ is an exact equality below fourth order. 
As a consequence, the values of $\hat{O}[\mathcal{T}_{\mathbf{S}}(T)]$ from different thermal lines are narrowly distributed around the mean value, suggesting the use of Monte Carlo sampling over thermal lines to estimate vibrational averages. This is accomplished by choosing the elements of $\mathbf{S}$ at random for each sampling point.

Two different versions of the method are discussed, depending on whether the odd term cancellation introduced in Eq.~(\ref{eq:odd_term_suppression}) is exploited or not. The first scheme is:
\begin{equation}
\langle \hat{O}(T)\rangle_{\mathrm{TL}}=\frac{1}{n}\sum_{i=1}^{n}\hat{O}[\mathcal{T}_{\mathbf{S}_i}(T)]. \label{eq:tl}
\end{equation}
The second scheme is a simple extension to include opposite pairs of thermal lines:
\begin{equation}
\langle \hat{O}(T)\rangle_{\mathrm{TL}_2}=\frac{1}{n}\sum_{i=1}^{n}\frac{1}{2}\left(\hat{O}[\mathcal{T}_{\mathbf{S}_i}]+\hat{O}[\mathcal{T}_{-\mathbf{S}_i}]\right). \label{eq:tl2}
\end{equation}
Equations~(\ref{eq:tl}) and (\ref{eq:tl2}) are the second central result of this paper. In the following, Monte Carlo sampling over the vibrational wave function in Eq.~(\ref{eq:mc}) is termed WF, Monte Carlo sampling over thermal lines in Eq.~(\ref{eq:tl}) is termed TL, and Monte Carlo sampling over opposite pairs of thermal lines in Eq.~(\ref{eq:tl2}) is termed TL$_2$.

It is worth describing at this stage what is accomplished by Monte Carlo sampling over thermal lines. It follows from $\langle \hat{O}(T)\rangle\simeq\hat{O}[\mathcal{T}_{\mathbf{S}}(T)]$ that the property distribution arising from thermal lines is narrower than the distribution arising from the wave function, and TL and TL$_2$ require fewer sampling points than WF to achieve the same statistical uncertainty. 
Therefore, the computational cost cross-over with the quadratic method will occur at smaller system sizes in TL and TL$_2$, making random sampling over thermal lines a promising approach to calculate quantum and thermal averages at small computational cost. 

Regarding the accuracy of the method, it should be noted that it is exact if the expansion of the property of interest is exactly quadratic. Otherwise, the method is subject to an uncontrolled bias affecting the evaluation of the expectation value of interest. Nonetheless, although the derivation of thermal lines was based on the quadratic approximation, the normal mode coordinates appearing in Eq.~(\ref{eq:thermal_lines}) have large amplitudes, and all vibrational modes contribute for any given configuration. As a consequence, terms beyond quadratic order contribute when using thermal lines, and, as demonstrated numerically in Sec.~\ref{sec:results} for a few systems and physical properties, multi-phonon terms are largely captured by thermal lines. 

\section{Computational details} \label{sec:comput}

Calculations of diamond, silicon, and L-alanine (C$_3$H$_7$NO$_2$) are reported below. 
All calculations were performed using the plane-wave pseudopotential DFT code {\sc castep}~\cite{castep} and ultrasoft ``on the fly'' pseudopotentials to describe the ionic cores~\cite{PhysRevB.41.7892}. The exchange-correlation energy was described within the local density approximation~\cite{PhysRevLett.45.566,PhysRevB.23.5048} for diamond and silicon, and within the generalized gradient approximation~\cite{PhysRevLett.77.3865} for L-alanine. Energy cut-offs and electronic BZ grids were chosen to reduce the energy differences between different distortions of the crystal structures below $1$~meV/atom. 

All structures were relaxed to reduce the forces on atoms below $1$~meV/\AA, and the stress on the simulation cell below $0.1$~GPa. The resulting structures for diamond and silicon have lattice parameters $a=3.529$~\AA\@ and $a=5.394$~\AA\@, respectively. The L-alanine crystal structure, containing $4$ molecules in the primitive cell, is the same as that used in Ref.~\onlinecite{monserrat_nmr_tdep_original}, with orthorhombic symmetry and lattice parameters $a=5.806$~\AA, $b=5.940$~\AA, and $c=12.274$~\AA. 

The harmonic frequencies and eigenvectors were calculated using the finite displacement method~\cite{phonon_finite_displacement}, averaging over positive and negative displacements of amplitude $0.005$~\AA. The resulting matrix of force constants was Fourier transformed to obtain the dynamical matrices at points $\mathbf{k}$ in the vibrational BZ, and these were diagonalized to calculate the harmonic frequencies and eigenvectors.

For diamond and silicon, the vibrational correction to the thermal band gap was calculated by considering the valence band maximum at $\Gamma$, and the conduction band minimum along the symmetry line $\Gamma$-$X$. For L-alanine, the chemical shielding tensor was calculated with the GIPAW theory~\cite{gipaw,gipaw_ultrasoft} as implemented in {\sc castep}~\cite{castep}.

\section{Results} \label{sec:results}

\subsection{Energy}

\begin{figure}
\centering
\includegraphics[scale=0.40]{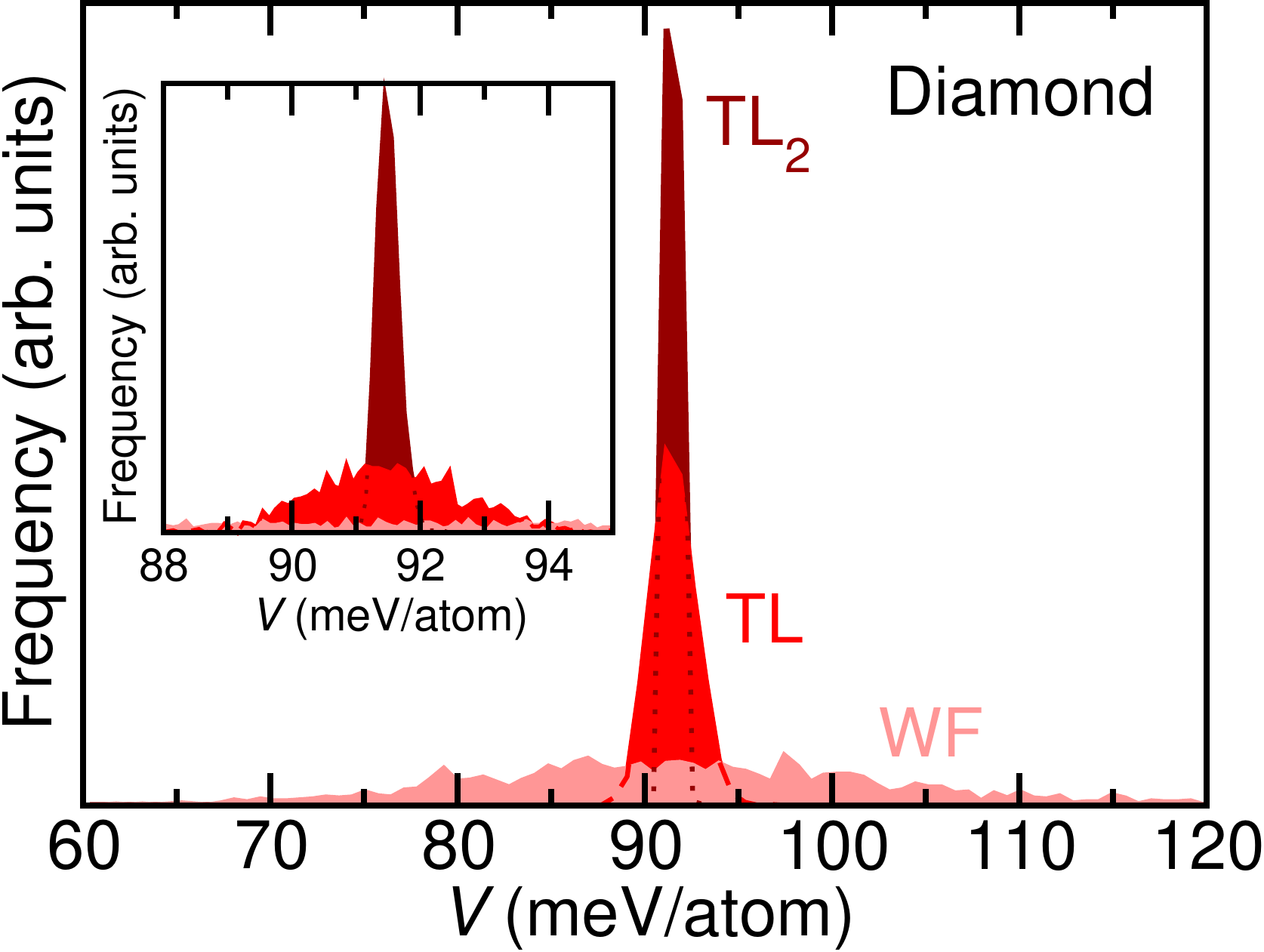}
\includegraphics[scale=0.40]{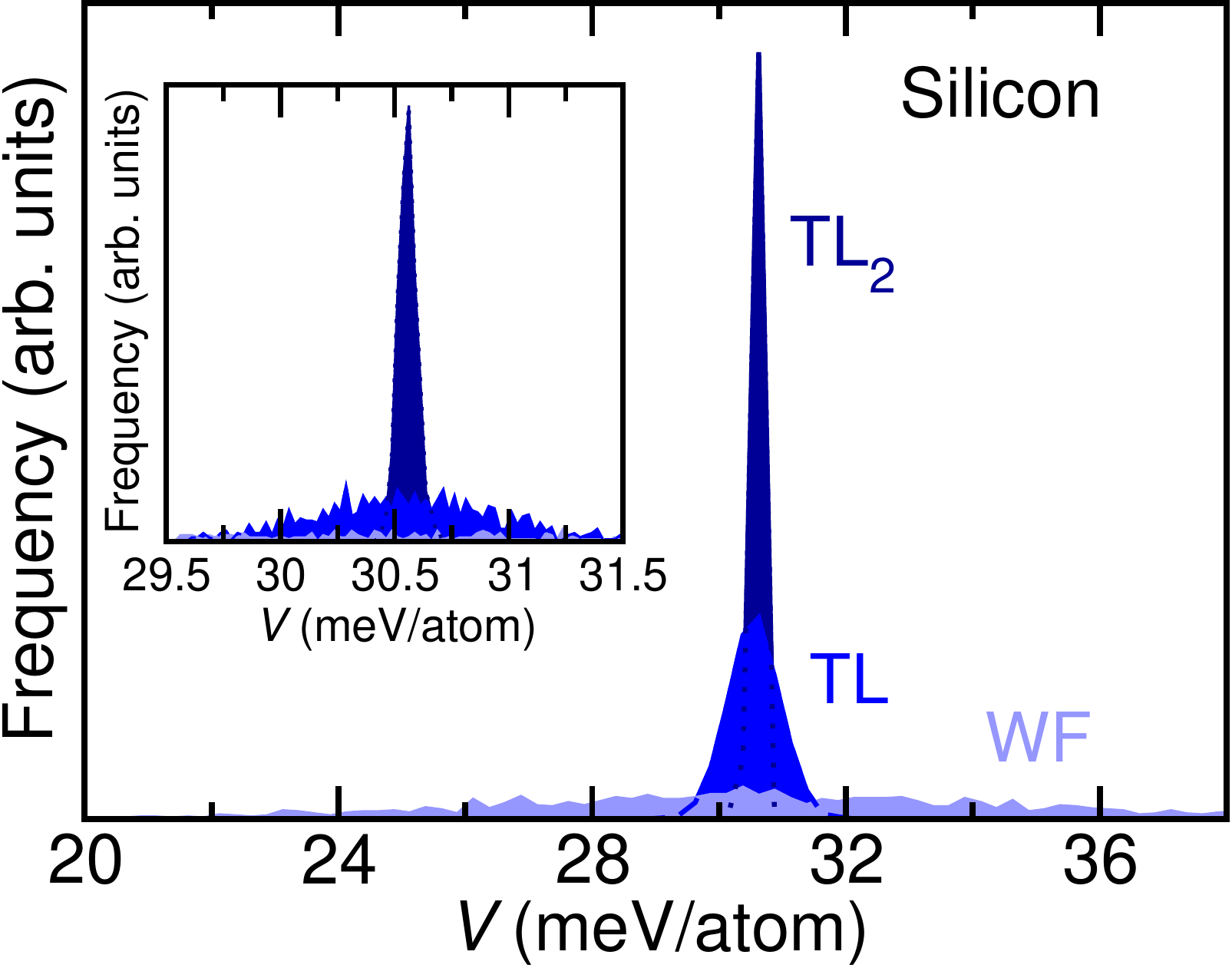}
\caption{Potential energy $V$ distributions arising from WF, TL, and TL$_2$ samplings for diamond and silicon. The insets show the same data but with a narrower energy window about the mean. The calculations have been performed using $54$-atom simulation cells.}
\label{fig:energy}
\end{figure}

The potential energy $V$ is the first observable of interest. In Fig.~\ref{fig:energy}, the zero-temperature distributions of energies obtained using WF, TL, and TL$_2$ sampling are shown for diamond and silicon. The insets show a narrower energy window about the average, to appreciate the width of the distributions arising from TL and TL$_2$. The energy distributions show the clear improvement achieved by sampling over thermal lines rather than over the full vibrational wave function. In particular, TL$_2$ sampling has a $5\sigma_V$ range of less than $1$~meV/atom in diamond, and less than $0.25$~meV/atom in silicon.
As expected, the potential energy obtained by means of the quadratic approximation in Eq.~(\ref{eq:quad_tdep}) is equal to the value obtained with the WF, TL, and TL$_2$ samplings.

The approximation in Eq.~(\ref{eq:quad_approx}) is equivalent to the harmonic approximation for the potential energy. The harmonic approximation is known to work very well for diamond and silicon, and the narrow distributions arising from TL and TL$_2$ are therefore to be expected. The finite width of the distributions arises from small anharmonic vibrational energy terms.


The vibrational frequencies $\omega_{n\mathbf{k}}$ and normal mode coordinates $q_{n\mathbf{k}}$ are required to construct the sampling points in WF, TL, and TL$_2$ methods. These are typically determined using finite displacements or density functional perturbation theory in first-principles calculations, both of which deliver the vibrational harmonic free energy (as a sum or integral over vibrational frequencies). Therefore, WF, TL, and TL$_2$ are not advantageous for calculating vibrational free energies within DFT, and the main purpose of the above discussion was to exemplify the methods proposed in this paper. Thermal lines could still be useful for vibrational free energy calculations using methods beyond DFT, such as QMC: one could define the thermal lines using DFT, and then calculate the QMC energies along them to obtain approximate QMC free energies. The exploration of this possibility is left for a future study.

\subsection{Electronic band gap}

\begin{figure}
\centering
\includegraphics[scale=0.40]{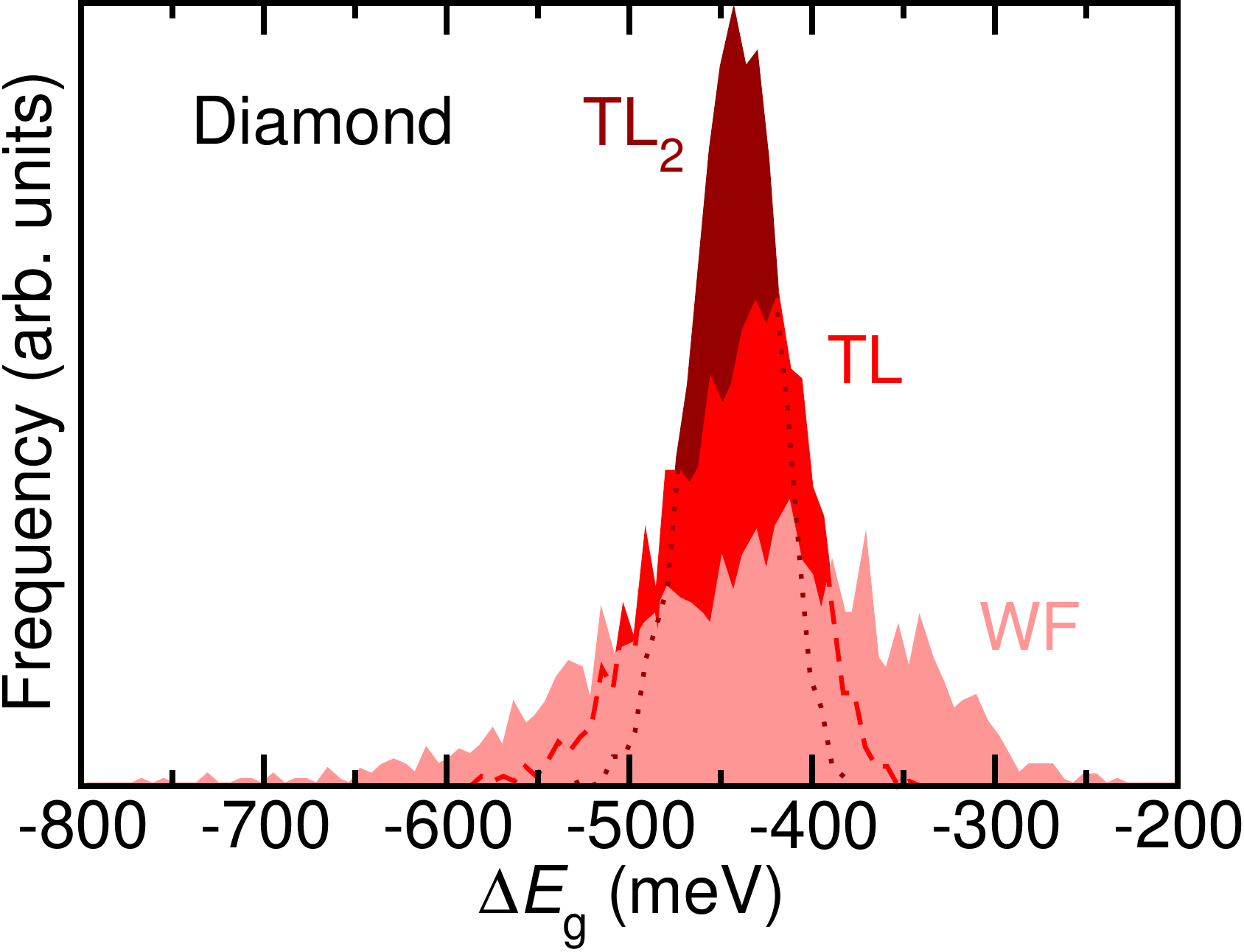}
\includegraphics[scale=0.40]{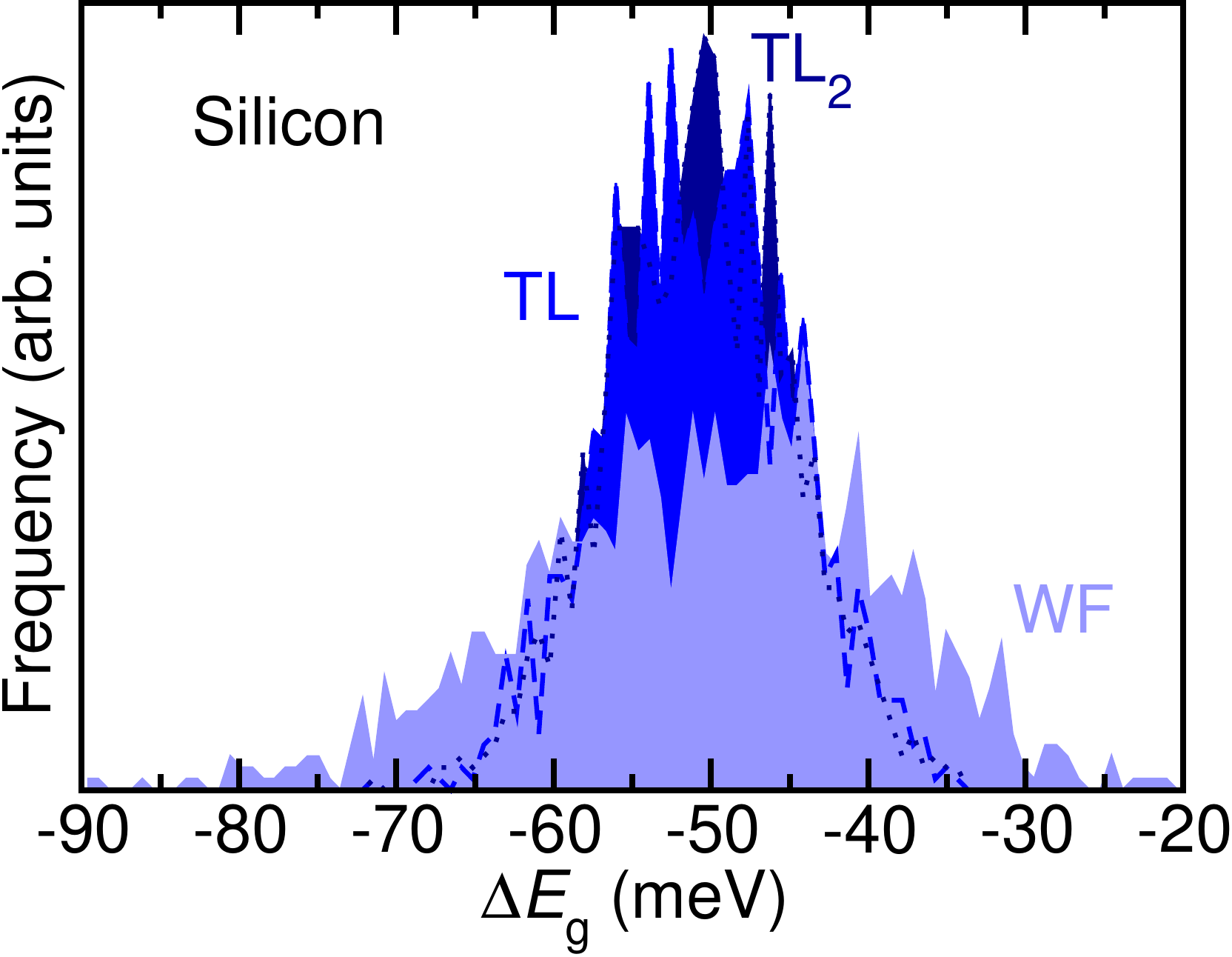}
\caption{ZP band gap correction $\Delta E_{\mathrm{g}}$ distributions arising from WF, TL, and TL$_2$ samplings for diamond and silicon. The calculations have been performed using $54$-atom simulation cells.}
\label{fig:gap}
\end{figure}

The electronic thermal band gap $E_{\mathrm{g}}$ is the second observable of interest, and the reported quantity is the band gap correction, $\Delta E_{\mathrm{g}}=\langle E_{\mathrm{g}}\rangle - E_{\mathrm{g}}(\mathbf{0})$, the difference between the vibrationally averaged gap and the static lattice gap. 
The distributions of zero-temperature band gap corrections obtained using WF, TL, and TL$_2$ for diamond and silicon are shown in Fig.~\ref{fig:gap}. As already observed for the energy, the widths of the gap distributions obtained sampling thermal lines are narrower than the width of the distribution obtained sampling the full vibrational wave function, although the difference is not as dramatic for the gap. This is to be expected, as there is no equivalent to the harmonic approximation for the band gap to make Eq.~(\ref{eq:quad_approx}) an equality.

\begin{figure}
\centering
\includegraphics[scale=0.40]{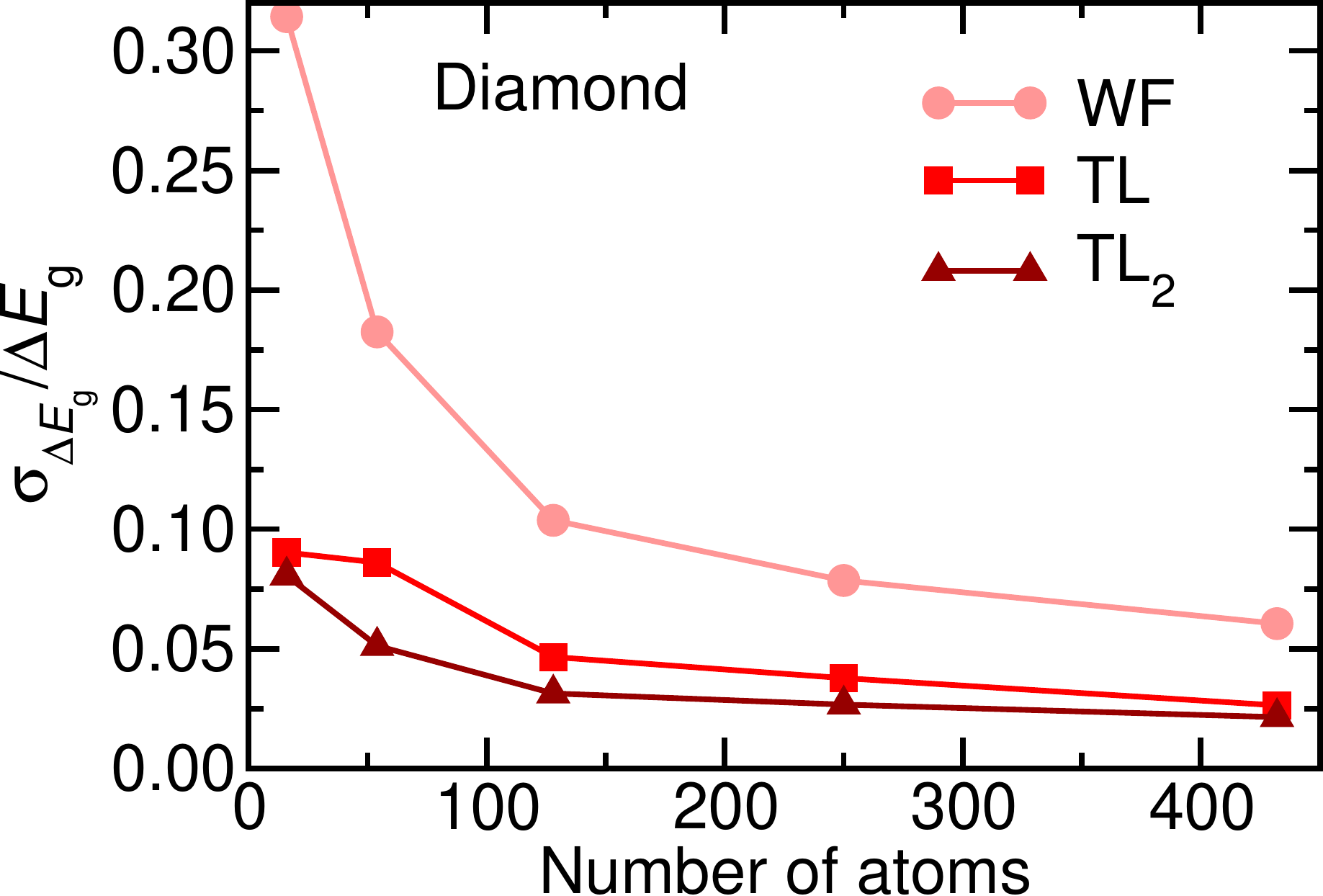}
\includegraphics[scale=0.40]{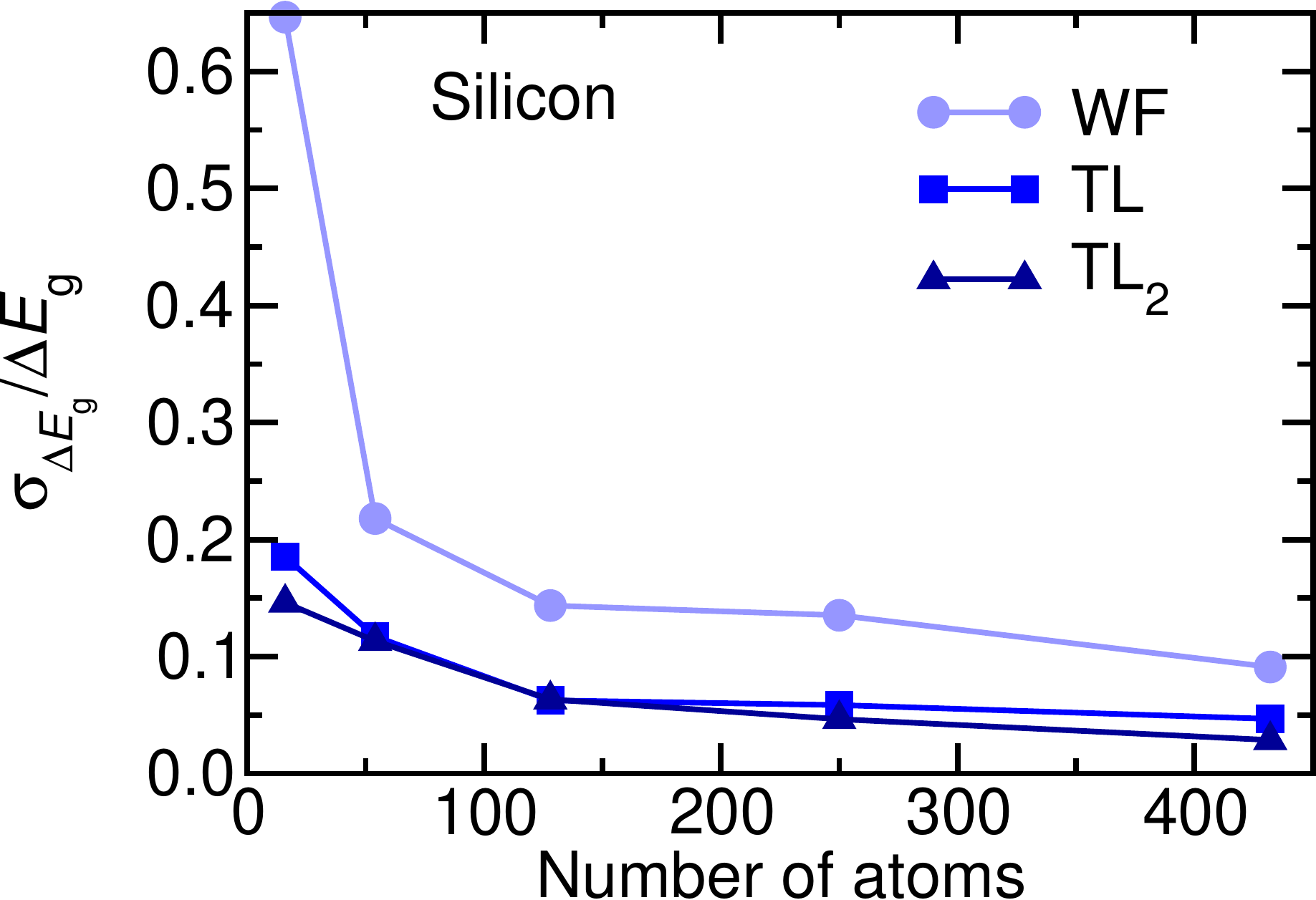}
\caption{Relative standard deviation of the ZP band gap corrections of diamond and silicon as a function of system size, comparing WF, TL, and TL$_2$ sampling methods.}
\label{fig:variance}
\end{figure}

The standard deviation of the ZP band gap correction distributions does not increase with system size, in fact, it slightly decreases with system size for diamond and silicon, as shown in Fig.~\ref{fig:variance}. Taking diamond as an example, sampling over thermal lines (both TL and TL$_2$) leads to relative uncertainties $\sigma_{\Delta E_{\mathrm{g}}}/\Delta E_{\mathrm{g}}$ below $0.05$ for simulation cells containing $128$ atoms or more, and for $432$-atom cells, the relative uncertainty is halved to $0.025$. 
The results therefore suggest that accurate vibrational corrections to electronic band gaps can be calculated using a small number of sampling points, irrespective of system size. This opens the door to routine calculations to investigate systems containing many atoms.

It is interesting to note that sampling over thermal lines (both TL and TL$_2$) delivers a result that is in better agreement with accurate Monte Carlo sampling of the wave function than the results obtained using the quadratic approximation (see Table~\ref{tab:mc_vs_quad}). This might seem surprising as the mathematical derivation of thermal lines is based on the same expansion used to define the quadratic method. However, the configurations associated with thermal lines have contributions from all normal modes, and at relatively large amplitudes $\sqrt{\langle q_{n\mathbf{k}}^2\rangle}$, unlike those used in the quadratic method. Therefore, they contain information about the higher-order terms in Eq.~(\ref{eq:expansion}), and the results in Table~\ref{tab:mc_vs_quad} show that this leads to more accurate expectation values than the quadratic method. The results in Table~\ref{tab:mc_vs_quad} show that the contribution of high-order terms in diamond ranges from $3$ to $9$\% depending on system size. Similar contributions are found in silicon, in agreement with the results of Ref.~\onlinecite{patrick_molecule_solid_2014}.

To further confirm that thermal lines accurately capture higher order terms missing in the quadratic approximation, the ZP correction to the band gap of the molecular crystal NH$_3$ is considered. This quantity has recently been shown to exhibit a strong non-quadratic behaviour, with large contributions from multi-phonon terms~\cite{molec_crystals_elph}. The space group of NH$_3$ is $P2_13$, and the primitive cell contains four molecules. The calculations have been performed using {\sc castep}~\cite{castep} and the PBE functional~\cite{PhysRevLett.77.3865} corrected with the TS scheme to describe dispersion interactions~\cite{ts_vdW}, with the same numerical parameters as those in Ref.~\onlinecite{molec_crystals_elph}. For a primitive cell ($\Gamma$-phonon coupling only), the ZP corrections to the gap are $-0.60$~eV, $-0.61$~eV, and $-0.60$~eV, using MC, TL, and TL$_2$ sampling, respectively. 
These results numerically confirm the accuracy of calculations based on thermal lines for a highly non-quadratic case. The standard deviation of the distributions of ZP corrections are $0.23$~eV, $0.20$~eV, and $0.14$~eV, for MC, TL, TL$_2$ sampling respectively. For this strongly non-quadratic example, TL$_2$ significantly improves upon TL, an observation that is also made in Sec.~\ref{subsec:nmr} for the chemical shielding tensor of L-alanine, with a dominant linear component in the expansion of Eq.~(\ref{eq:expansion}). 

\begin{table*}
\setlength{\tabcolsep}{10pt} 
\caption{Comparison of the ZP band gap corrections of diamond and silicon, obtained using WF, TL, and TL$_2$ sampling. The results from Ref.~\onlinecite{monserrat_elph_diamond_silicon} obtained using the quadratic approximation with the same computational parameters as those used in this study are also included.} 
\label{tab:mc_vs_quad}
\begin{tabular}{cccccc}
\hline
\hline
System  & BZ grid &  WF (meV)  & TL (meV)  & TL$_2$ (meV) &  Quadratic~\cite{monserrat_elph_diamond_silicon} (meV)   \\
\hline
Diamond & $3\times3\times3$ & $-439$ & $-444$ & $-443$ & $-401$ \\
        & $4\times4\times4$ & $-317$ & $-317$ & $-315$ & $-292$ \\
        & $5\times5\times5$ & $-347$ & $-347$ & $-347$ & $-325$  \\
        & $6\times6\times6$ & $-344$ & $-344$ & $-343$ & $-334$  \\ 
\\
Silicon & $3\times3\times3$ & $-51$ & $-51$ & $-51$ & $-53$ \\
        & $4\times4\times4$ & $-52$ & $-52$ & $-52$ & $-52$ \\
        & $5\times5\times5$ & $-62$ & $-62$ & $-62$ & $-60$  \\
        & $6\times6\times6$ & $-58$ & $-59$ & $-59$ & --  \\
\hline
\hline
\end{tabular}
\end{table*}

\begin{figure}
\centering
\includegraphics[scale=0.40]{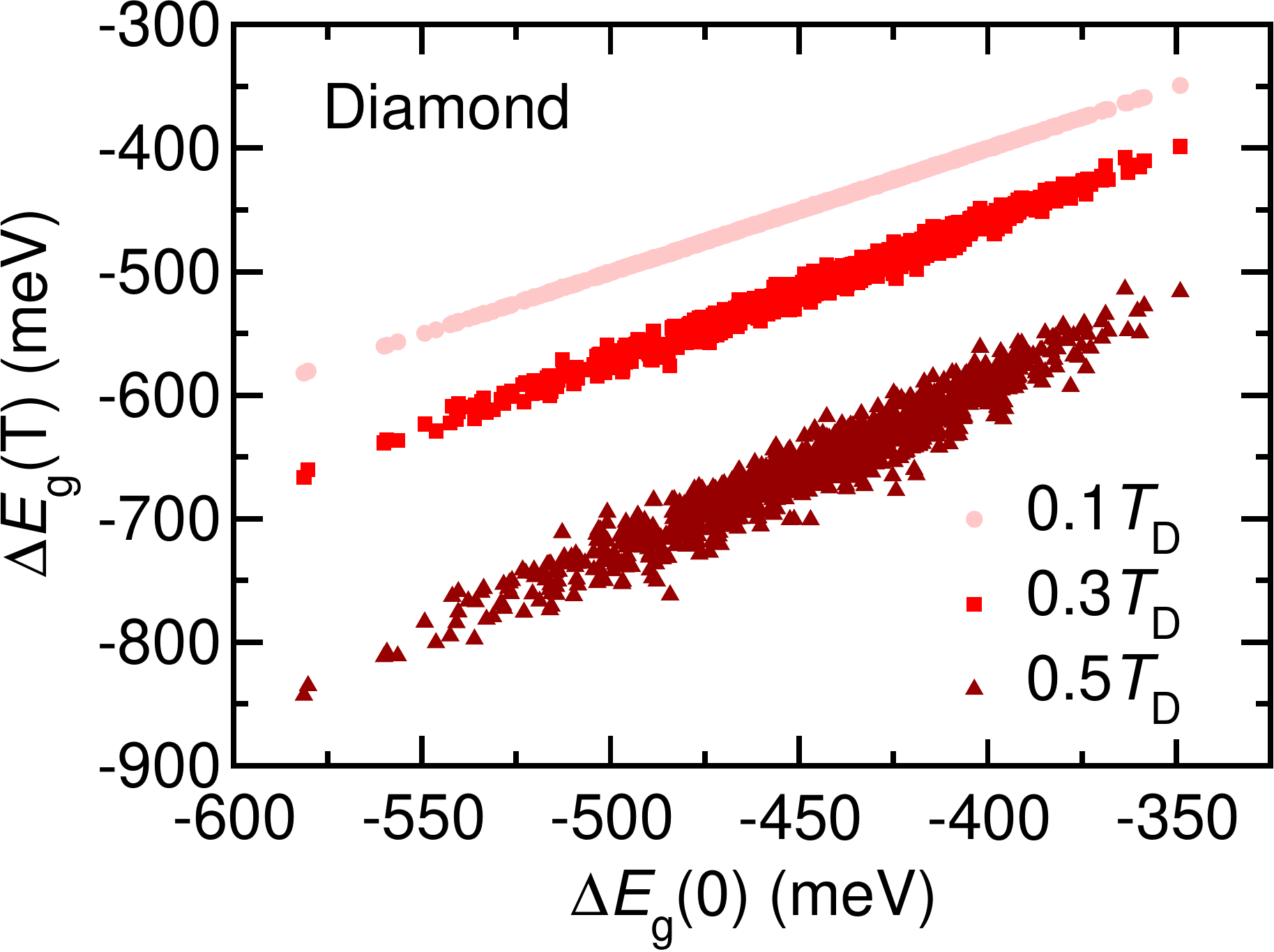}
\includegraphics[scale=0.40]{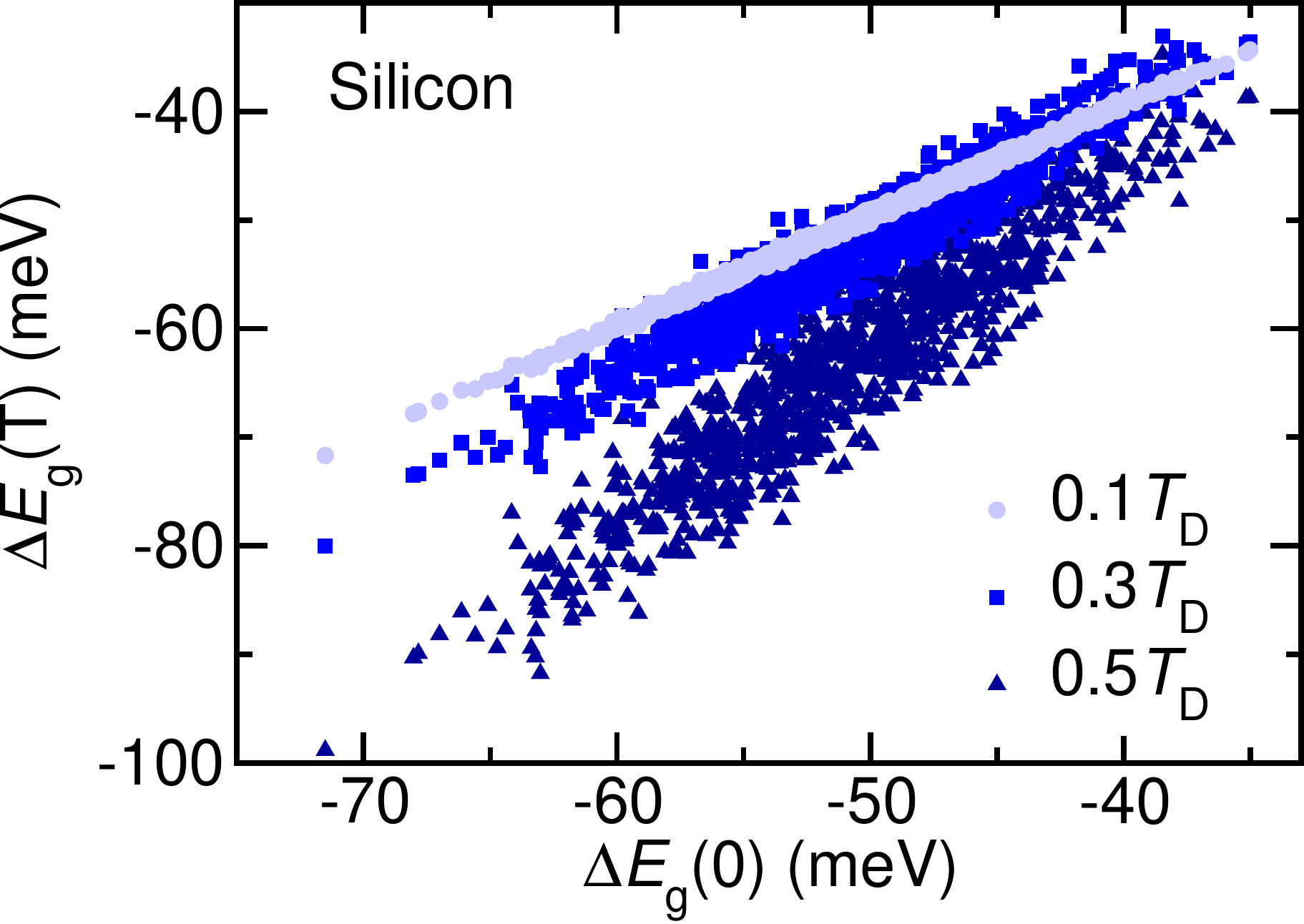}
\caption{Band gap corrections at temperatures of $0.1T_{\mathrm{D}}$, $0.3T_{\mathrm{D}}$, and $0.5T_{\mathrm{D}}$ against the ZP band gap correction. For diamond, $T_{\mathrm{D}}=2230$~K, and for silicon, $T_{\mathrm{D}}=645$~K. The results correspond to $54$-atom simulation cells.} 
\label{fig:tdep}
\end{figure}

\subsection{Temperature dependence}

The results described above correspond to quantum averages using the quantum point $T=0$. The same conclusions would be reached with calculations at other temperatures $T>0$, but using instead configurations associated with the points $T$ along the thermal lines.

In this section, a simple approach to exploit thermal lines to calculate the temperature dependence of a quantity of interest is described, and the vibrational correction to the band gap is used as an example. Let $\mathcal{T}_{\overline{\mathbf{S}}}$ be the \textit{mean thermal line}, defined as the thermal line on which the property of interest has a value closest to the vibrational average: 
\begin{equation}
\overline{\mathbf{S}}(T)=\argmin_{\mathbf{S}}|\hat{O}[\mathcal{T}_{\mathbf{S}}(T)]-\langle \hat{O}(T)\rangle_{\mathrm{TL/TL}_2}|. \label{eq:mean}
\end{equation}
The minimization in Eq.~(\ref{eq:mean}) runs over the thermal lines $\mathcal{T}_{\mathbf{S}}$ sampled when calculating the vibrational average using TL or TL$_2$ at temperature $T$. 

After determining the mean thermal line at temperature $T$ using Eq.~(\ref{eq:mean}), the vibrational average of the property of interest at a different temperature $T'$ can be calculated by evaluating the property at point $T'$ along the mean thermal line, namely $\hat{O}[\mathcal{T}_{\overline{\mathbf{S}}(T)}(T')]$. The ability of the mean thermal line to accurately capture the temperature dependence is demonstrated in Fig.~\ref{fig:tdep} for diamond and silicon. The band gap correction at temperature $T$ is plotted against the ZP band gap correction, for a set of thermal lines. The strong correlation between the two sets of vibrational averages indicates that the use of the mean thermal line accurately captures temperature dependences. 

The mean thermal line determined according to Eq.~(\ref{eq:mean}) should be largely independent of the temperature at which TL or TL$_2$ sampling is performed, due to the correlation observed in Fig.~\ref{fig:tdep}. In Table~\ref{tab:temp}, thermal averages for diamond and silicon calculated using TL sampling are compared to those calculated using the mean thermal line, chosen using Eq.~(\ref{eq:mean}) at a range of temperatures. The results confirm the weak temperature dependence of the definition of the mean thermal line, but also indicate that sampling in the middle of the temperature range of interest leads to the best results. This is to be expected as the correlation in Fig.~\ref{fig:tdep} decreases for increasing temperature difference. The largest error in diamond appears if the thermal line is determined at $T=0$~K, and then used to calculate the thermal average at $0.5T_{\mathrm{D}}$ ($T=1,115$~K, where $T_{\mathrm{D}}$ is the Debye temperature), and even in this case the error is smaller than $4$\%. For silicon, the largest error arises when choosing the mean thermal line at $0.5T_{\mathrm{D}}$ ($T=322$~K), and then calculating the thermal average at $T=0$~K, with an error of $5$\%.  For both diamond and silicon, choosing the mean thermal line with TL or TL$_2$ sampling at a central temperature of $0.1T_{\mathrm{D}}$ or $0.3T_{\mathrm{D}}$ leads to the best results overall.

\begin{table*}
\setlength{\tabcolsep}{10pt} 
\caption{Comparison of the thermal averages of the ZP band gap corrections of diamond and silicon, obtained using TL sampling over thermal lines, and using the mean thermal line $\mathcal{T}_{\overline{\mathbf{S}}}$. The mean thermal line is determined using Eq.~(\ref{eq:mean}) at a range of temperatures. The results correspond to $54$-atom simulation cells, and the TL averages include $1,000$ points at each temperature.}
\label{tab:temp}
\begin{tabular}{cccccc}
\hline
\hline
System & Method  &  $0$~K (meV)   & $0.1T_{\mathrm{D}}$ (meV) & $0.3T_{\mathrm{D}}$ (meV) &  $0.5T_{\mathrm{D}}$ (meV)  \\
\hline
Diamond & TL                                                & $-444$ & $-444$ & $-509$ & $-653$  \\
& $\mathcal{T}_{\overline{\mathbf{S}}(0)}$                  & $-444$ & $-444$ & $-500$ & $-631$ \\ 
& $\mathcal{T}_{\overline{\mathbf{S}}(0.1T_{\mathrm{D}})}$  & $-444$ & $-444$ & $-513$ & $-661$ \\ 
& $\mathcal{T}_{\overline{\mathbf{S}}(0.3T_{\mathrm{D}})}$  & $-443$ & $-443$ & $-509$ & $-653$ \\ 
& $\mathcal{T}_{\overline{\mathbf{S}}(0.5T_{\mathrm{D}})}$  & $-440$ & $-440$ & $-506$ & $-653$ \\ 
\\
Silicon & TL                                                & $-50.9$ & $-50.4$ & $-52.7$ & $-63.1$  \\
& $\mathcal{T}_{\overline{\mathbf{S}}(0)}$                  & $-50.9$ & $-50.4$ & $-52.1$ & $-61.7$ \\ 
& $\mathcal{T}_{\overline{\mathbf{S}}(0.1T_{\mathrm{D}})}$  & $-50.7$ & $-50.4$ & $-52.9$ & $-63.3$ \\ 
& $\mathcal{T}_{\overline{\mathbf{S}}(0.3T_{\mathrm{D}})}$  & $-51.0$ & $-50.6$ & $-52.7$ & $-62.6$ \\ 
& $\mathcal{T}_{\overline{\mathbf{S}}(0.5T_{\mathrm{D}})}$  & $-53.3$ & $-52.6$ & $-53.4$ & $-63.1$ \\ 
\hline
\hline
\end{tabular}
\end{table*}

The mean thermal line results suggest that a single point per temperature might be sufficient to calculate accurate thermal averages. 


\subsection{Chemical shielding} \label{subsec:nmr}

The nuclear magnetic resonance chemical shielding tensor $\bm{\tau}$ is considered as the property of interest in this section, and vibrational averages of the isotropic chemical shift
\begin{equation}
\tau_{\textrm{iso}}=\frac{1}{3}\mathrm{tr} \bm{\tau}, 
\end{equation}
are reported. Again, the correction arising from atomic vibrations will be the quantity of interest, $\Delta\tau_{\mathrm{iso}}=\langle\tau_{\mathrm{iso}}\rangle-\tau_{\mathrm{iso}}(\mathbf{0})$.

\begin{figure}
\centering
\includegraphics[scale=0.80]{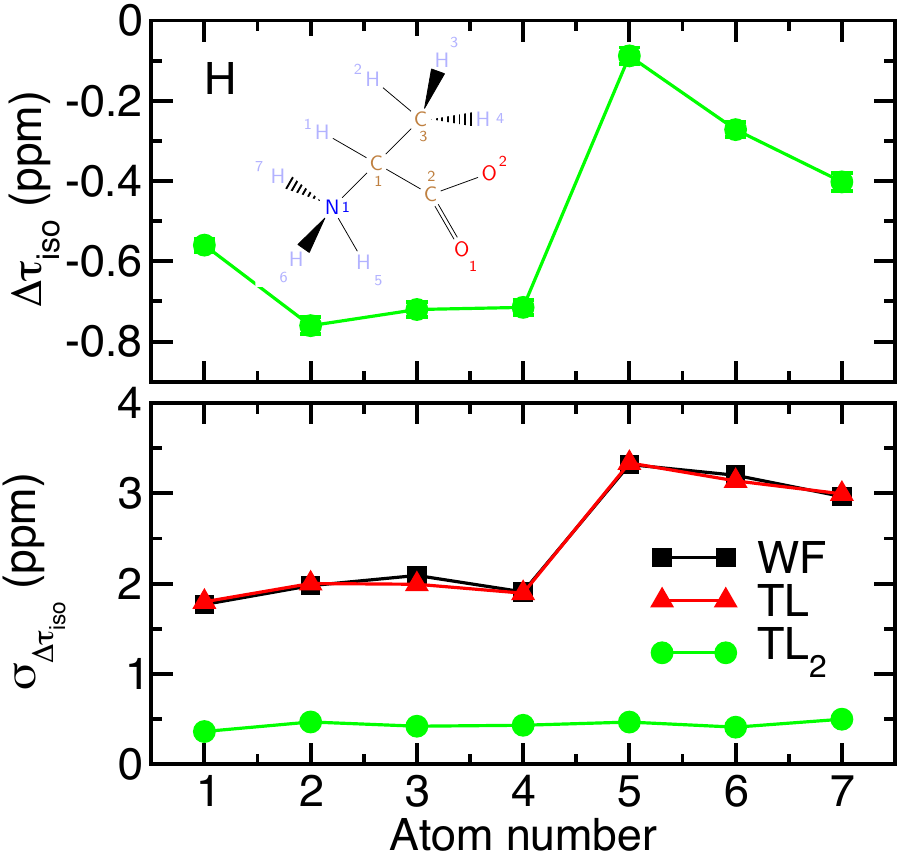}
\caption{ZP correction to the isotropic chemical shift $\Delta\bm{\tau}_{\mathrm{iso}}$ of the hydrogen atoms in L-alanine (top) and the corresponding standard deviations $\sigma_{\Delta\bm{\tau}_{\mathrm{iso}}}$ of the distributions arising from WF, TL and TL$_2$ sampling (bottom). The inset shows the atom numbering for the L-alanine molecule. The solid lines linking the data points are a guide to the eye only.}
\label{fig:nmr_h}
\end{figure}

\begin{figure}
\centering
\includegraphics[scale=0.40]{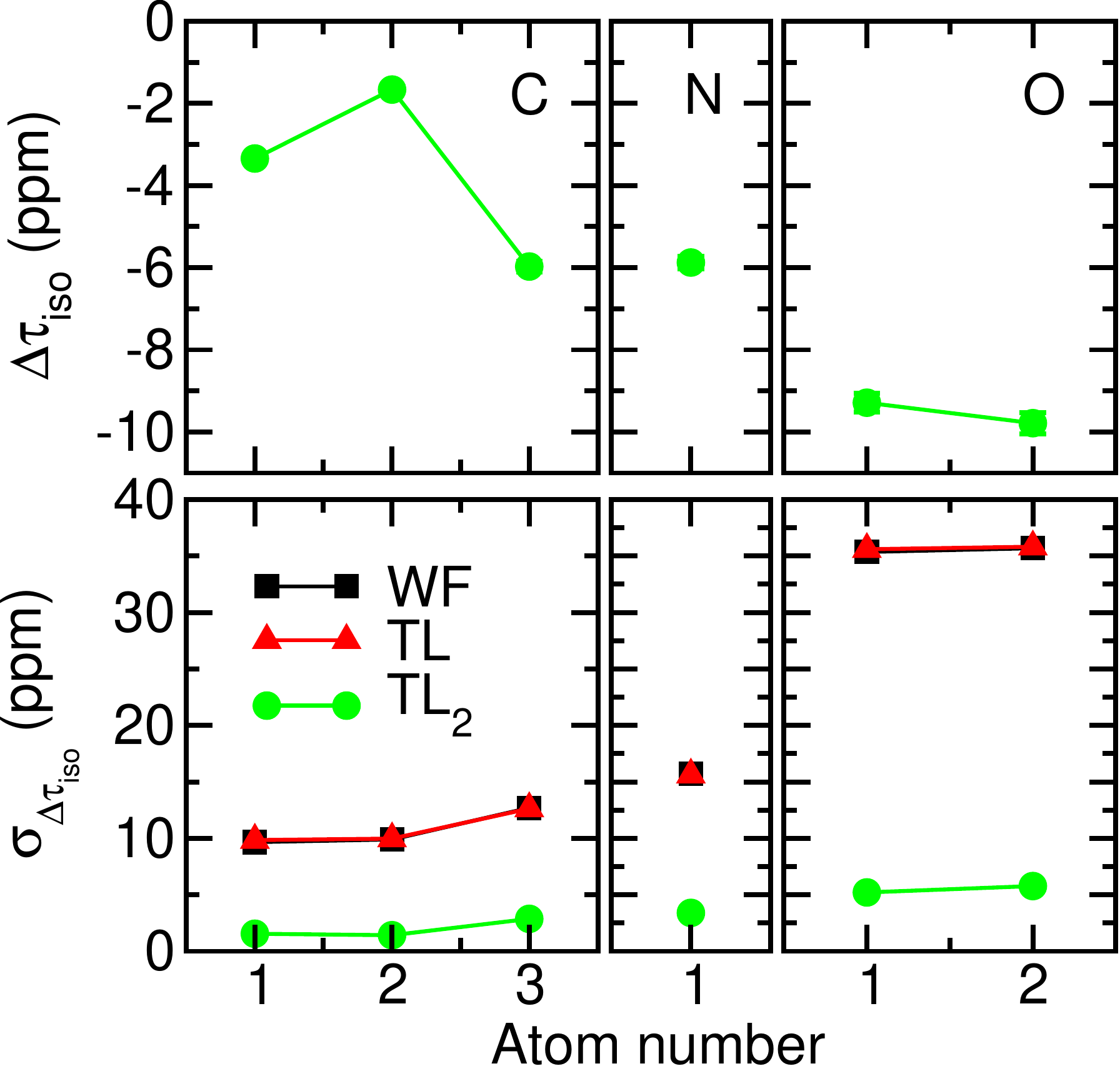}
\caption{ZP correction to the isotropic chemical shift $\Delta\bm{\tau}_{\mathrm{iso}}$ of the carbon, nitrogen, and oxygen atoms in L-alanine (top) and the corresponding standard deviations $\sigma_{\Delta\bm{\tau}_{\mathrm{iso}}}$ of the distributions arising from WF, TL and TL$_2$ sampling (bottom). The atom numbering is that shown in Fig.~\ref{fig:nmr_h}. The solid lines linking the data points are a guide to the eye only.}
\label{fig:nmr_cno}
\end{figure}

Figures~\ref{fig:nmr_h} and \ref{fig:nmr_cno} show the ZP correction to the isotropic chemical shift for all atoms and species in the L-alanine molecule in the crystalline phase, calculated using $2\times500$ sampling points following TL$_2$ sampling (top diagrams). The atom numbering is shown in the inset of Fig.~\ref{fig:nmr_h} (note that the hydrogen atoms attached to the carbon and nitrogen were mislabeled in Ref.~\onlinecite{monserrat_nmr_tdep_original}). The corrections are in good agreement with those reported in Ref.~\onlinecite{monserrat_nmr_tdep_original} and obtained using the WF and quadratic methods. 

In the bottom diagrams of Figs.~\ref{fig:nmr_h} and~\ref{fig:nmr_cno}, the standard deviation of the isotropic chemical shift, $\sigma_{\Delta\bm{\tau}_{\mathrm{iso}}}$,  is shown for each atom and species corresponding to WF, TL, and TL$_2$ sampling. The static lattice isotropic chemical shifts are listed in Table~\ref{tab:nmr}. The behaviour of the standard deviations is qualitatively different to the examples of the energy and bang gap of diamond and silicon considered above. For the isotropic chemical shifts, the use of TL sampling does not reduce the standard deviation compared to WF sampling. The use of TL$_2$ sampling reduces the standard deviation by an order of magnitude, significantly more than the corresponding reduction found for the energy and band gap. This behaviour can be rationalized by noting that the change of the components of the chemical shielding tensor with increasing normal mode amplitude is predominantly linear rather than quadratic. The mean value of the vibrational average is moved away from the thermal lines, and therefore sampling over thermal lines is not better than sampling over the full vibrational wave function (note that it is not worse either). By contrast, averaging over pairs of opposite thermal lines in TL$_2$ sampling exactly removes the linear component, and as a consequence the standard deviation of the distribution is dramatically reduced. Therefore, for properties with large odd components in the expansion of Eq.~(\ref{eq:expansion}), the extension of sampling over thermal lines to include pairs of opposite thermal lines provides a significant advantage.

\begin{table}
\setlength{\tabcolsep}{10pt} 
\caption{Static lattice isotropic chemical shifts for the L-alanine molecular crystal. The atom numbering is shown in Fig.~\ref{fig:nmr_h}.}
\label{tab:nmr}
\begin{tabular}{ccc}
\hline
\hline
Species  & Atom number &  $\bm{\tau}_{\mathrm{iso}}(\mathbf{0})$ (ppm) \\
\hline
H & $1$ & $26.87$ \\
H & $2$ & $29.65$ \\
H & $3$ & $29.17$ \\
H & $4$ & $29.73$ \\
H & $5$ & $19.03$ \\
H & $6$ & $21.72$ \\
H & $7$ & $23.22$ \\
\hline
C & $1$ & $120.25$ \\
C & $2$ & $-13.10$ \\
C & $3$ & $152.75$ \\
\hline
N & $1$ & $181.01$ \\
\hline
O & $1$ & $-27.43$ \\
O & $2$ & $-47.98$ \\
\hline
\hline
\end{tabular}
\end{table}

The quadratic method was compared to WF sampling in Ref.~\onlinecite{monserrat_nmr_tdep_original}, and the former was proposed as a computationally inexpensive approach for the inclusion of thermal effects in NMR, given the smaller number of sampling points that had to be considered. For L-alanine, a total of $306$ sampling points are required to calculate the coupling of the chemical shielding tensor to $\Gamma$-point phonons using the quadratic method~\cite{monserrat_nmr_tdep_original}. Using $306$ sampling points with the WF method leads to statistical uncertainties in the ZP correction to the isotropic chemical shifts in the range $0.10$--$0.20$~ppm in hydrogen, $0.55$--$0.70$~ppm in carbon, $0.90$~ppm in nitrogen, and $2.03$~ppm in oxygen. These statistical uncertainties are too large for many applications: experimental chemical shifts of hydrogen and carbon atoms are commonly reported with accuracies of $0.1$~ppm~\cite{bmaltose_exp_nmr}. For comparison, using $306$ points with TL$_2$ sampling leads to statistical uncertainties in the vibrational corrections to the isotropic shifts that are significantly smaller than those of WF sampling; they are $0.02$--$0.03$~ppm, $0.15$~ppm, $0.28$~ppm, and $0.45$~ppm in hydrogen, carbon, nitrogen, and oxygen, respectively. 

TL$_2$ sampling therefore requires a similar number of sampling points to the quadratic method for primitive cell calculations of vibrational renormalizations in L-alanine. For calculations using larger simulation cells, TL$_2$ should be the method of choice.

\section{Conclusions} \label{sec:conclusions}

I have introduced thermal lines to effectively explore the vibrational phase space of solids. This is accomplished because the value of a physical property for an atomic configuration corresponding to point $T$ along a thermal line is approximately equal to the thermal average of that property at temperature $T$. Monte Carlo sampling over thermal lines can be used to calculate accurate vibrational averages using a small and size-independent number of sampling points.

The use of thermal lines is demonstrated by calculating quantum and thermal averages of the potential energy and the electronic band gaps of diamond and silicon, and of the chemical shielding tensor of L-alanine. In all cases, the use of thermal lines leads to accurate results that are obtained using a small number of sampling points. 

Thermal lines should be useful for calculating quantum and thermal averages of structural, optical, electronic, and magnetic properties in an accurate and simple manner. Future work will focus on demonstrating the wide applicability of thermal lines for calculations of vibrational averages of various physical properties, using methods beyond semi-local DFT, and for systems containing a large number of atoms.

\acknowledgments

The author thanks M. Crispin-Ortuzar and J.H. Lloyd-Williams for helpful comments on the manuscript, and Robinson College, Cambridge, and the Cambridge Philosophical Society for a Henslow Research Fellowship.


\bibliography{/Users/bartomeumonserrat/Documents/research/papers/references/anharmonic.bib,/Users/bartomeumonserrat/Documents/research/papers/references/nmr.bib}

\end{document}